\documentclass[12]{article}
\usepackage{graphics,graphicx,amsbsy,amssymb,color}

\newcommand{\uu}{{\boldmath \mbox{$u$}}}

\newcommand{\rr}{{\boldmath \mbox{$r$}}}

\newcommand{\qq}{{\boldmath \mbox{$q$}}}

\newcommand{\vv}{{\boldmath \mbox{$v$}}}

\newlength{\defbaselineskip}
\newcommand{\setlinespacing}[1]%
           {\setlength{\baselineskip}{#1 \defbaselineskip}}

\title{\textbf{
TEMPERATURE}}
\author{Miroslav Grmela \footnote{ e-mail:
miroslav.grmela@polymtl.ca}\\
\'{E}cole Polytechnique  Montr\'{e}al,
  C.P.6079 suc. Centre-ville,\\
 Montr\'{e}al, H3C 3A7,  Qu\'{e}bec, Canada}

 \date{}

\begin{document}

\maketitle

\begin{abstract}

Conversations about     weather, environment, health, cuisine, and even politics, all involve the word "temperature". It was an attempt to understand the working of heat engines   that gave the temperature a  clear definition.  In this Note we put the equilibrium-thermodynamics definition into a larger context of the multiscale thermodynamics, the multiscale rate thermodynamics, and nonphysical environments.

\end{abstract}

KEYWORDS: Equilibrium and Nonequilibrium Thermodynamics and Statistical Mechanics

MSC: 82C03

\section{Equilibrium temperatures}\label{Sec1}

The  fundamental stone  on which the equilibrium thermodynamics stays is the observation of the existence of the approach of externally unforced macroscopic systems to  equilibrium states at which no time evolution takes place (called 0-th law of thermodynamics). All systems considered in this section are either at equilibrium or are approaching equilibrium. More general situations are considered in Sections \ref{Sec2} and \ref{Sec3}.

\subsection{Macroscopic equilibrium temperature}\label{Sec1.1}

Macroscopic systems are characterized in the classical equilibrium thermodynamics by an internal energy per unit volume  $\epsilon$ and the number of moles per unit volume $\nu$. In this Note we limit ourselves only to the energy. The number of moles remains unchanged and we omit it in our notation.

Macroscopic systems are surrounded by two types of walls. One allows free passing of $\epsilon$ (Wall $\|$) and the other prevents it (Wall $\sharp$). Let us consider a particular situation in which a macroscopic system that is surrounded by Wall $\sharp$ is  divided into two subsystems ($\mathcal{S}_1, \mathcal{S}_2$) by Wall $\|$. The system is not at equilibrium at the time $t=0$.  The internal energies of the two subsystems  are $\epsilon_1$ and $\epsilon_2$. According to 0-th law of thermodynamics the system starts to evolve  in time and reaches eventually an equilibrium state. The equilibrium thermodynamics addresses only the final outcome of the time evolution but not the time evolution itself (see \ref{Sec1.3} where such time evolution is addressed). The state reached as $t\rightarrow \infty$ is the state at which entropy $s(\epsilon)$ reaches its maximum (MaxEnt principle). The individual nature of the macroscopic system under investigation is expressed in equilibrium thermodynamics in the relation  $s=s(\epsilon)$ (called a fundamental thermodynamic relation). All  relations  $s=s(\epsilon)$ are required to be  real valued, sufficiently regular, and concave function. Moreover, the entropy $s$ as well as the internal energy $\epsilon$ are assume to be extensive state variables  (i.e. $\epsilon=\epsilon_1 +\epsilon_2$ and $s(\epsilon_1,\epsilon_2)=s(\epsilon_1)+s(\epsilon_2)$).

With these requirements  the MaxEnt principle implies
\begin{equation}\label{eq0}
\frac{ds}{d\epsilon_1}=\frac{ds}{d\epsilon_2}
\end{equation}
This means that at the equilibrium state the temperatures $T_1$ and $T_2$ of the two subsystems $\mathcal{S}_1$ and $\mathcal{S}_2$, defined by
\begin{equation}\label{eq1}
\frac{1}{T}=\frac{ds(\epsilon)}{d\epsilon}=\epsilon^*>0
\end{equation}
are equal. The equilibration of temperatures in the approach to equilibrium  is also the  basis of their measurement. Let  $\mathcal{S}_2$ be a system in which states   are characterized by a collection $\xi$ of state variables. The state variables in  $\xi$  include the internal energy $\epsilon$ but they also include other state variables. For example if $\mathcal{S}_2$ is a human body then $\xi$ includes many biological state variables like for instance the variables characterizing states of nerves. Their coupling to the internal energy then brings about  the feeling of the temperature. This biological thermometer has always existed, has always been known, and moreover, it has been  absolutely essential for survival of humans and all living creatures. An objective measurement of the temperature is made by thermometers in which $\xi$ includes state variables, like for instance volume, that can be objectively quantified.

We mention two other important consequences of the above requirements on the fundamental thermodynamic relation $s=s(\epsilon)$ and on the internal energy $\epsilon$. First, we note that the extensibility of both $\epsilon$ and $s$ implies that $s$ is a 1-homogeneous function of $\epsilon$, i.e. $s=\frac{1}{\lambda}s(\lambda \epsilon)$ for $\lambda\in\mathbb{R}$. Consequently, $s=s(\epsilon)$ can be extended (by Euler relation) to $S=S(E,V)$; $S=E^*E+V^*V$, where $V$ is the volume, $S$ and $E$ is the entropy and the internal energy of the whole system (i.e. $S=Vs; E=V\epsilon$; $E^*=\epsilon^*=\frac{1}{T}; V^*=-\frac{P}{T}$ is the conjugate of $V$; $P$ has the physical interpretation of the pressure. The second consequence of the requirements on $s$ and $\epsilon$, namely that $\frac{ds(\epsilon)}{d\epsilon}=\epsilon^*>0$ is that there is a one-to-one relation between $s$ and $\epsilon$. We can consider $s$ to be a state variable and $\epsilon=\epsilon(s)$ as the fundamental thermodynamic relation.

Summing up, \textit{\textbf{the concept of temperature in the classical equilibrium thermodynamics requires: (i) a substance having an internal energy, (ii) walls allowing to pass or to stop passing it, (iii) the process of equilibration driven by the gradient of a potential called entropy (MaxEnt). The multiplicative inverse of the  temperature is the conjugate  $\epsilon^*$ (with respect to the entropy) of the internal energy; $\epsilon^*=\frac{1}{T}$.}} The MaxEnt principle makes the group of Legendre transformations  a fundamental group of thermodynamics and consequently the contact geometry in the three-dimensional space with coordinates $(s,\epsilon,\epsilon^*)$ is the natural geometry of thermodynamics. The Legendre transformations preserve the contact 1-form $ds-\epsilon^*d\epsilon$. The fundamental thermodynamic relation $s=s(\epsilon)$ makes (in the contact geometry formulation of the classical equilibrium thermodynamics) appearance as the Legendre submanifold (i.e. the two-dimensional manifold on which $ds-\epsilon^*d\epsilon=0$). The Legendre transformation  of $s=s(\epsilon)$ is $s^*=s^*(\epsilon^*)=\phi(\hat{\epsilon}(\epsilon^*),\epsilon^*)$ where
\begin{equation}\label{L1}
\phi(\epsilon,\epsilon^*)=-s(\epsilon)+\epsilon^*\epsilon
\end{equation}
and $\hat{\epsilon}(\epsilon^*)$ is a solution of $\frac{\partial \phi}{\partial \epsilon}=0$.

\subsection{Microscopic equilibrium temperature}\label{Sec1.2}

In this section we see macroscopic systems as  composed of $n \sim 10^{23}$ particles obeying (we assume) the classical mechanics.
State variables of the particles are $(r,v)\in \mathbb{R}^{6n}$ where $r$ are position coordinates and $v$ momenta of the particles. In the Liouville lift \cite{Liouville} the state variables are $x=f(r,v), f:\mathbb{R}^{6n}\rightarrow \mathbb{R}^+$. We shall denote the Liouville state space with the symbol $M$; $f(r,v)\in M$.

As in the classical thermodynamics we do not specify the time evolution describing the approach to equilibrium (0-th law of thermodynamics) but address only its final outcome by the MaxEnt principle. The entropy in this completely microscopic viewpoint of macroscopic systems is the Gibbs entropy $s(x)=-\int dr \int dv f(r,v)\ln f(r,v)$ (we put  the Boltzmann constant $k_B=1$) that is universal for all macroscopic systems. The individual nature of macroscopic systems is expressed in the energy
$e(x)=\int dr \int dv f(r,v)h(r,v)$, where $h(r,v)$ is the microscopic Hamiltonian of the particles. There is no internal energy $\epsilon$ since all the details of the dynamics are expressed in terms of $x$. There is no substance of unknown origin that  contributes  to the total energy $e$. The MaxEnt passage from $(s(x),e(x))$ to the fundamental thermodynamic relation $(s^*=s^*(\epsilon^*))$ is made by the Legendre transformation: $s^*=s^*(\epsilon^*)=\phi(\hat{x}(\epsilon^*),\epsilon^*)$, where
\begin{equation}\label{L2}
\phi(x,\epsilon^*)=-s(x)+\epsilon^* e(x)
\end{equation}
and $\hat{x}(\epsilon^*)$ is a solution to $\frac{\partial\phi}{\partial x}=0$: $\hat{f}(r,v)=\exp{-\epsilon^*h(r,v)}$.

We note that the temperature enters the Gibbs microscopic thermodynamics in the MaxEnt passage to the equilibrium (as the Lagrange multiplier in the thermodynamic potential (\ref{L2})) and then, when the passage is completed,  in  the equilibrium submanifold
\begin{equation}\label{Meq}
\mathcal{M}_{eq}=\{x\in M| \frac{\partial\phi(x,\epsilon^*)}{\partial x}=0\}
\end{equation}
representing the classical equilibrium thermodynamics in the Gibbs theory. The equilibrium manifold (\ref{Meq}) is in fact a family of manifolds parametrized by the temperature. We could now systematically  look at all possible geometrical characterization of the manifolds (\ref{Meq}) and investigate their possible physical (equilibrium-thermodynamic) interpretations. Results of such investigation would give  physical interpretations of the temperature and  offer also possible ways to measure it. Many views of the temperature are developed and reviewed in \cite{JouT}.
For instance, the most well known is the relation between the temperature  and  the average kinetic energy  ($\int dv \frac{v^2}{2m}\hat{f}(r,v)$)   in the case when $h(r,v)= \frac{v^2}{2m}$ (i.e. in the case when  the macroscopic system under investigation is an ideal gas, $m$ is the mass of one particle). If the macroscopic system under investigation is not an ideal gas (i.e. if $h(\rr,\vv)$ is not only the kinetic energy) then the average kinetic energy of the microscopic particles composing the macroscopic system (calculated for example in numerical simulations) is neither  directly related to the equilibrium manifold (\ref{Meq}) nor  to the temperature measured by standard thermometers.

Summing up, the temperature on the completely microscopic level of description does not exists. It emerges in the MaxEnt passage to the equilibrium state. It is a parameter in  the equilibrium manifolds (\ref{Meq}) expressing the classical equilibrium thermodynamics in the Gibbs microscopic thermodynamics.

\subsection{Mesoscopic equilibrium temperatures}\label{Sec1.3}

In this section we take a mesoscopic view of macroscopic systems. The state variable $x$ does not take into account  all the microscopic details as the  n-particle distribution function $f(r,v)$ does.  For instance $x$ may be   1-particle distribution function. With such state variable there are obviously many microscopic details that cannot be expressed.  The mesoscopic thermodynamics with 1-particle distribution function as state variable, called kinetic theory, has been pioneered by Ludwig Boltzmann \cite{Boltzmann}. The approach to equilibrium is  made in his theory again by MaxEnt but  the MaxEnt passage is not postulated but is made by following solutions of the Boltzmann  equation. This means that  the 0-th postulate of the equilibrium thermodynamics and the MaxEnt principle (postulated in both the classical equilibrium thermodynamics and the Gibbs microscopic equilibrium thermodynamics) appear in solutions of the Boltzmann equation.  The only problem with  the Boltzmann theory is its limited applicability. It addresses only one macroscopic system, namely the ideal gas.

How can the Boltzmann theory be extended to general macroscopic systems? One way to achieve it is to extract from the theory an abstract mathematical structure that is common also to other well established (i.e. tested with experimental observations) mesoscopic dynamical theories (in particular hydrodynamics). The original Boltzmann theory could be then seen as a particular realization of such abstract Boltzmann theory and other mesoscopic dynamical theories addressing general macroscopic systems as its other particular realizations. This path has been taken in \cite{Gr84}, \cite{GrPhys}, \cite{GO}, \cite{OG}. A similar path, but started with the plasma dynamics rather than  the Boltzmann dynamics, has been taken in \cite{MorrPla}, \cite{Kaufman}. Both investigations have been motivated by Arnold's formulation \cite{Arnold} of the Euler hydrodynamic equation as a noncanonical Hamilton equation. Arnold's investigation   introduced  geometry into mesoscopic dynamics.

The time evolution governed by the abstract Boltzmann equation
\begin{equation}\label{generic}
\dot{x}=\left[L(x)x^* -\Xi_{x^*}(x,x^*)\right]_{x^*=\phi_x(x,\epsilon^*)}
\end{equation}
describes indeed  the MaxEnt approach to  equilibrium states $\hat{x}(\epsilon^*)$  (that are solutions to $\phi_x(x,\epsilon^*)$ - (see \cite{book}). We use in (\ref{generic}) the following notation: $\dot{x}=\frac{\partial x}{\partial t}$; $\phi_x$ is a shorthand notation for $\frac{\partial\phi}{\partial x}$; $\phi(x)$ is the thermodynamic potential (\ref{L2}); $L(x)$ is a Poisson bivector expressing kinematics of the mesoscopic state variable $x$; $\Xi(x,x^*)$ is a dissipation potential. We refer the reader to \cite{book} where all properties that $L$ and $\Xi$ are required to satisfy and many their particular  realizations and physical interpretations can be found. The Boltzmann kinetic equation as well as for instance the Navier-Stokes-Fourier hydrodynamic equations
are  indeed  particular realizations of (\ref{generic}).

Another feature of Eq.(\ref{generic})  shows  its belonging to thermodynamics.  Not only the passage from an initial state to the  asymptotically (i.e. $t\rightarrow\infty$) reached state (i.e. to the equilibrium state $\hat{x}(\epsilon^*)$) is made by a Legendre transformation (by MaxEnt)) but the complete time evolution leading to $\hat{x}(\epsilon^*)$ can be displayed as a sequence of infinitesimal Legendre transformations (transformations preserving the contact structure). In the  lift of (\ref{generic}) (see \cite{Grlift}, \cite{Ess1}, \cite{Ess2}), \cite{GrRT}) from the space with coordinates $x$ to the space with coordinates $(x,x^*,\phi)$ that is equipped with the contact 1-form $d\phi-x^*dx$ the time evolution governed by (\ref{generic}) takes place on a Legendre submanifold (the submanifold on which $d\phi-x^*dx =0$ ) and preserves $d\phi-x^*dx$. As we noted already in Section \ref{Sec1.1}, the fundamental group of thermodynamics on all levels is the group of Legendre transformations. The lift (see \cite{Grlift}, \cite{Ess1}, \cite{Ess2}), \cite{GrRT}) carries dynamics from mechanics (in which the fundamental group is the group of canonical transformation preserving the symplectic structure) to thermodynamics (in which the fundamental group is the group of Legendre transformations preserving the contact structure).

Still another way to see (\ref{generic}) is as a reversal of the extension of equilibrium thermodynamics to thermodynamic with fluctuations (see e.g. \cite{Callen}). The MaxEnt reduction made by following solutions of (\ref{generic}) is a top-down passage from levels that take into account more details to the level of equilibrium thermodynamics. The extension of equilibrium thermodynamics to thermodynamic with fluctuations is a bottom-up extension from the level of equilibrium thermodynamics to a level  involving more details (involving fluctuations). Both the top-down and the bottom-up passages are driven by the entropy.

The state variable $x$  in this section is a mesoscopic state variable. This means that there are details in dynamics that cannot be expressed with it. The contribution of the ignored details  to the total energy  is  an internal energy $\epsilon$. The way we take it into account in mesoscopic dynamical theories gives the physical interpretation to the temperature. In the rest of this section we shall discuss two ways to include the internal energy $\epsilon$: In Section \ref{Sec1.3.1} by modifying the thermodynamic potential $\phi(x)$, and in Section \ref{Sec1.3.2} by adopting the internal energy $\epsilon$ into the set of state variables.

\subsubsection{Modifications of the  energy and the entropy}\label{Sec1.3.1}

The classical equilibrium thermodynamics (Section \ref{Sec1.1}) and the Gibbs microscopic thermodynamics (Section \ref{Sec1.2}) are two extreme answers to the question of how to take into account the microscopic nature of the macroscopic systems. In the former in a single state variable (in the internal energy $x=\epsilon$), in the latter by taking into account all the mechanical details in the n-particle distribution function $x=f(r,v)$. Taking the mesoscopic point of view of macroscopic systems we have to attempt to combine the two extremes.

First we do not modify the chosen mesoscopic state variable $x$ but turn to modifications of the energy $e(x)$. Let $x$ be 1-particle distribution function $f(r,v)$. If the macroscopic system is an ideal gas (as in the Boltzmann investigation) then $e(x)=\int dr\int dv\frac{v^2}{2m}$ is the  energy that can be completely expressed in terms of $x$. Any particle-particle interactions need at least a 2-particle distribution function to express it. In the case of long range interactions we can approximate the two particle distribution function $f(r_1,v_1,r_2,v_2)$ by $f(r_1,v_1)f(r_2,v_2)$. The energy $e(x)$ with such approximation is then the Vlasov mean field energy.

There are two other ingenious ways to express complex interactions with the 1-particle distribution functions. Both extend the 1-particle distribution function $f(r,v)$ to $f(r,v,\zeta)$ and  both
address macroscopic systems composed of complex molecules.

Let the molecules be long chain-like macromolecules. Due to their snake like shapes they move, in the presence of other macromolecules, like snakes. De Gennes \cite{deGenn} expressed this type motion by choosing $\zeta=(R,\chi)$, where
$\chi\in\mathbb{R}$ is a coordinate along the backbone of the chain and $R\in\mathbb{R}^3$ is the tangent vector at the backbone coordinate $\chi$. The snake like motion, called reptation by de Gennes, is mathematically expressed as diffusion along the backbone.

A similar type of extension has been introduced independently by Ruggeri and Sugyiama \cite{Rugg} for fluid composed of complex macromolecules with an internal energy inside them. In their extension $f(r,v)\rightarrow f(r,v,\zeta)$, the new coordinate $\zeta\in\mathbb{R}$ is the internal energy of a single macromolecule.

Still another way to introduce the  influence of ignored mechanical details is to modify the entropy $s(x)$. We have seen in Section \ref{Sec1.2} that on the completely microscopic level where all mechanical details are taken into account and the entropy $s(x)$ is universal (the Gibbs entropy) for all systems. Its role is to sweep away  unimportant details and reveal the level of the equilibrium thermodynamics. The same role plays the entropy (the Boltzmann entropy) in the Boltzmann kinetic theory (except that the sweeping is made by following solutions of the Botzmann equation) since the kinetic energy of the particles composing an ideal gas composed of point particles is the complete energy.

Let us replace the ideal gas with the van der Waals gas. The particles composing it interact via long range attractive forces and short range hard-core repulsive forces. Following van Kampen \cite{vanKamp} we still choose the 1-particle disstribution function $f(r,v)$  as the mesoscopic state variable. The long range attractive force is expressed in van Kampen's theory as the Vlasov mean field force, but the hard-core repulsive force is expressed in the modification of the entropy $s(x)$. From the point of view of the classical equilibrium thermodynamics, the hard core interactions are thus expressed as a heat.
The Boltzmann entropy $-\int dr\int dv f(r,v)\ln f(r,v)$ is replaced by $-\int dr\int dv f(r,v)\ln f(r,v)+ \int dr\int dv f(r,v)(1-b\int dv f(r,v)) $ where $b$ is the volume of the particle. In this modification of the Boltzmann theory the entropy ceases thus its universality. It is taking the role of contributing to the specification of the individual nature of the van der Waals fluid. We recall that on the level of the equilibrium thermodynamics the individual nature of macroscopic systems is expressed only in the entropy, the internal energy is universal (it serves as a state variable).

All the ways to express the influence of  details (that cannot be expressed in terms of the chosen mesoscopic state variable $x$) that we discussed above  modify the equilibrium manifold (\ref{Meq}). Consequently, they also modify the physical interpretation of the temperature $\epsilon^*=\frac{1}{T}$ parametrizing them.

\subsubsection{Internal energy as an extra state variable}\label{Sec1.3.2}

The second family of attempts to take into account effect of details that cannot be expressed in terms of the chosen mesoscopic state variable $x$ follow the example of fluid mechanics. We recall the formulation of fluid mechanics. Macroscopic systems (fluids) are seen as a continuum \cite{Euler}. Its states are characterized by fields (real valued functions of the position coordinate $r$) of mass density $\rho(r)$ and the momentum $u(r)$. The time evolution equation is the Euler equation (that is the Newton equation for the continuum) supplemented by the local mass conservation equation. How shall we express the internal energy?
Following the example of the equilibrium thermodynamics we simply adopt it as an additional state variable except now it is a field rather than a real number as it is in the equilibrium thermodynamics. We assume that the extra forces that arise due to the macroscopic motion expressed in the momentum field $\uu(\rr)$ do not influence the local equilibrium thermodynamics. The 0-th law of thermodynamics still holds locally.
The energy $e(x)=\int dr \frac{u^2(r)}{2\rho(r)} +\int dr\epsilon(r)$ is the complete energy. This setting supplemented by the local energy conservation equation and the Navier-Stokes-Fourier dissipation is the first and still the most used and the most successful mesoscopic dynamical theory. The temperature in fluid mechanics is  defined as in the classical equilibrium thermodynamics by (\ref{eq1}) except that it is a local temperature (it is a field) since both
the entropy $s$ and the internal energy are local.

A need for an extension of the classical fluid mechanics arises for instance in fluids involving long range  (on the macroscopic scale) interactions \cite{Hill}, \cite{Syros}, in the heat conduction on the microscopic scale \cite{Jou}, \cite{Van}, \cite{Kov} and in complex fluids (as for example polymeric fluids and suspensions \cite{book}). All extensions are made by adopting extra fields, we denote them by the symbol $\xi(\rr)$,  into the set of the state variables; $x=(\rho(\rr),\uu(\rr),\xi(\rr),\epsilon^{(\xi)}(\rr))$. The total energy per unit volume $e=\frac{1}{2}\int d\rr \frac{u^2(r)}{2\rho(r)} + \frac{1}{2}\int d\rr e^{(\xi)}(\xi(\rr),\rr) + \frac{1}{2}\int d\rr \epsilon^{(\xi)}(\rr)$.
where $e^{(\xi)}(\xi(\rr),\rr)$ is the energy that can be expressed in terms of the adopted new fields.

There are  two ways to see macroscopic systems with $x=(\rho(\rr),\uu(\rr),\xi(\rr),\epsilon^{(\xi)}(\rr))$ chosen to characterized their states. First, we see them as in Sections \ref{Sec1.2}, \ref{Sec1.3}, \ref{Sec1.3.1} and second as in fluid mechanics. In the first view the macroscopic systems approach to the  equilibrium manifold (\ref{Meq}). Individual nature of macroscopic systems is expressed  the energy  $e(x)=\frac{1}{2}\int d\rr \frac{u^2(r)}{2\rho(r)} + \frac{1}{2}\int d\rr e^{(\xi)}(\xi(\rr),\rr) + \frac{1}{2}\int d\rr \epsilon^{(\xi)}(\rr)$ and the entropy $s(x)=\frac{1}{V}\int d\rr s^{(\xi)}(\rho(\rr),\uu(\rr),\xi(\rr),\epsilon^{(\xi)}(\rr))$.  In the second view, we see macroscopic systems in the same way as in fluid mechanics except that  we include into the fields that are under our control (in the classical fluid mechanics such  fields are $(\rho(\rr), \uu(\rr))$ ) also the extra fields $\xi(\rr)$. For example, let  one of the fields in $\xi(\rr)$ be heat flux $\qq(\rr)$.  We fix it. The heat flux then plays the role of an external force that prevents approach to equilibrium (the 0-th law of thermodynamics ceases to hold) and consequently there is no entropy driving to the equilibrium (or to the local equilibrium as in fluid mechanics) and thus there is also no temperature (or local temperature) (\ref{eq1}). There is however still a rate temperature that we shall introduce in the next Section.

Before leaving this section we  make a  thought   experiment. We live on the scale of fluid mechanics, the world that we directly observe is the world that is well described by fluid mechanics. We imagine that we descend   gradually to smaller scales. Ultimately,  we  reach the scale of microscopic particles. On such scale we see around us  a permanent motion, there is no equilibrium and no temperature. Let us pose in our descend  to the microscopic scale at the scale of macromolecules. Instead of ourselves   we can also  imagine some kind of robots or biological machines (we shall call them MMR robots or MMR machines) operating on such  macromolecular scale. What we (or MMR robots) see is the physical behavior described with models that use $x=(\rho(\rr),\uu(\rr),\xi(\rr),\epsilon^{(\xi)}(\rr))$ as state variables   (provided $\xi(\rr)$ is composed of fields describing states of macromolecules). In general, there is no equilibrium, the 0-th law of thermodynamics does not hold and there is no temperature (\ref{eq1}).

\section{Rate temperatures}\label{Sec2}

The classical equilibrium thermodynamics and the classical fluid mechanics are  well established  theories. This means that solutions to governing equations of a more microscopic well established dynamical theory (as for example a kinetic theory)  has to display an approach to them.
Returning  to Boltzmann's pioneering investigation of the mesoscopic dynamics, we have already shown that the approach to the classical equilibrium thermodynamics is indeed displayed in the original as well as the abstract Boltzmann equation (\ref{generic}). We now turn to the approach to the classical fluid mechanics. Two paths  to investigate this problem have been suggested and explored. The first, taken by Chapman and Enskog \cite{ChE}, search an invariant (or approximately invariant) manifold on which  the 1-particle distribution function  $f(r,v)$ settles after  some time. The second, introduced by Grad \cite{Grad}, cast first the Boltzmann equation to a hierarchy (Grad hierarchy) and then closes it.  The second path has more the spirit of  thermodynamic, we shall therefore follow it.

The  0-th law of equilibrium thermodynamics  gave us the physical basis of the equilibrium thermodynamics. We shall call the existence of the approach of a well established and autonomous mesoscopic dynamical theory to another well established and autonomous dynamical theory that takes into account less details a 0-th law of rate thermodynamics. We use the name "rate thermodynamics" since in order to formulate thermodynamics we need an approach to a fixed point (to a state at which no time evolution takes place). This is indeed the case in the approach to equilibrium states but it is not the case in the approach to another dynamics. Such approach however turns out to be an approach to  fixed points if we consider the time evolution in the space of vector fields rather than the time evolution in the state space. The finally reached fixed points are the vector fields governing the time evolution in the reduced dynamical theory. The time evolution in the space of vector fields is called a rate time evolution. We thus call the thermodynamics arising in its investigation a rate thermodynamics.

The equation governing the time evolution in the space of vector fields is obtained by lifting Eq.(\ref{generic}) governing the time evolution in the state space to  cotangent (or tangent) bundles. The geometrical formulation of such lifts as well as their physical interpretation is discussed in detail in \cite{Ess1}, \cite{Ess2},\cite{GrRT}. The essence of the lifts is to consider in (\ref{generic}) the variable $x^*$ as an independent variable. We rewrite Eq.(\ref{generic}) into the form
\begin{equation}\label{generic1}
\dot{x^*}=\mathbb{G}\mathcal{R}_{x^*}(x,x^*)
\end{equation}
where
\begin{equation}\label{Rayl}
\mathcal{R}(x,x^*)=e^{\dag}<x^*,L(x)e_x(x)> -\Xi(x,x^*)
\end{equation}
is a Rayleighian, $\mathbb{G}=(\phi_{xx})^{-1}$, $e^{\dag}(x)=\frac{1}{T^{(rate)}(x)}$, $T^{(rate)}(x)$ is the rate temperature. The Poisson bivector $L$  as well as the dissipation potential $\Xi$ in (\ref{Rayl}) are  assumed to be degenerate in the sense that $Ls_x=0$  and $\Xi$ depends on $x^*$ only through its dependence on $Kx^*$, where $K$ is a linear operator satisfying $Ke_x=0$.

The Rayleighian plays in the rate thermodynamics the same role as the thermodynamic potential $\phi$ (see (\ref{L2})) plays in the equilibrium thermodynamics. The first term $<x^*,L(x)e_x(x)>$ is the rate energy (replacing the energy $e$ in (\ref{L2})) and the second term $\Xi(x,x^*)$ is the dissipation potential (replacing the entropy $s$ in (\ref{L2})). Note that the dissipation potential is closely related to the rate entropy (the rate entropy equals $<x^*,\Xi_{x^*}>$, note that in the particular case when $\Xi$ is a quadratic function of $x^*$ the entropy production equals twice the dissipation potential).

A few comments are in order. First, we note that the abstract Boltzmann  equation (\ref{generic}) in its reinterpretation as an equation (\ref{generic1}) governing the time evolution of $x^*$ rather than $x$ recovers and extends the Onsager principle. The reduced time evolution is obtained as an extremum (it can be shown a minimum, see  \cite{Ess1}, \cite{Ess2},\cite{GrRT}) of the Rayleighian. This is the statement of the Onsager principle \cite{Ray},\cite{OnP}, \cite{OM},\cite{Doi}. Equation
(\ref{generic1}) shows it and in addition it also shows how the reduced vector field is reached. Moreover, Eq.(\ref{generic1})  also provides the physical foundation of the Onsager principle (it is the 0-th law of the rate thermodynamics).

The second comment is about the presence of external forces. In the  equilibrium thermodynamics the external forces  prevent approach to equilibrium are thus excluded. They are not excluded in the rate thermodynamics. Dynamical theories reduce to dynamical theories involving less details also in the presence of external forces. The Rayleighian can include external forces.

In the third comment we note that the temperature $T^{(rate)}$ in the Rayleighian is, in general,  different from the equilibrium temperature $T$. First of all, we see that the rate temperature (as well as all quantities in the formulation of rate thermodynamics)  is independent of $x^*$ but it depends on $x$. For instance, let $x$ be hydrodynamic fields and $x^*$ are constitutive relations. The multiplicative inverse of the rate temperature  is the derivative of the rate entropy with respect to the rate energy. The 0-th law of rate thermodynamics on which the rate thermodynamics is based is in this case is the approach of constitutive relations to, say, Navier-Stokes-Fourier constitutive relations. The rate temperature, that addresses the approach, depends on the hydrodynamic fields. In the stationary (in the space of hydrodynamic fields) states the rate temperature is time independent but in general in nonstationary states it changes in time.

 Direct measurements of the rate temperature would need  Wall $\|$ freely passing the rate energy and and Wall $\sharp $  stopping its passing. We do not see  how to make such walls. More promising would be to observe fluctuations of a mesoscopic dynamical theory. We have noted in Section \ref{Sec1.1} that in the equilibrium thermodynamics the bottom-up  extension of the equilibrium theory to include fluctuations is a reversal process of the top-down reduction governed by (\ref{generic}). Both the extension and the reduction are driven by the entropy and both provide ways to measure the temperature.
We conjecture that similarly in the context of the rate thermodynamics a bottom-up extension of a mesoscopic dynamical theory (for instance hydrodynamics) to include fluctuations that is reversal process of the top-down reduction governed by (\ref{generic1}). Both the extension and the reduction we expect to be driven by the Rayleighian and both we expect to offer a way to measure the rate temperatures. We hope to investigate this conjecture more in the future and we also hope that we shall be joined by  interested readers.

\section{Nonstandard temperatures}\label{Sec3}

Everyday dealing with hot and cold and with the temperature  characterizing them   led to use the temperature  also as a measure of, for instance,  the intensity of  social interactions. In this section we discuss briefly two nonstandard temperatures.

\subsection{Granular temperature}\label{Sec.3.1}

Granular media, as for instance sand, are  macroscopic systems that  can be regarded in the same way as we regarded macroscopic systems in the previous sections. However, their particular structure and particular behavior turn our attention  to features that do not arise and are not studied in other macroscopic systems. For instance we are interested what triggers an avalanche (i.e. what triggers   granular systems with motionless granular particles  to start  to behave as  fluids).
With such interests in mind Sam Edwards \cite{Edwards} has introduced a "granular  temperature" that he calls fluffiness or also a compactivity.

As we pointed out in Section \ref{Sec1.1}  a quantity that plays the role of the temperature  in the classical thermodynamics needs (i) a medium having "an internal energy", (ii) walls through which the medium can freely pass or be completely stopped, (iii) a process of equilibration. Edwards has chosen the medium to be the space in the granular system that is not occupied by the grains. The  "internal energy" is its volume, and the equilibration the process is shaking.  The problem of developing this elegant idea into a multiscale theory of granular media has been found  challenging.

\subsection{Rumor temperature}\label{Sec3.2}

In this example we turn to politics. We indeed hear and read the word temperature in this context. For example we read: "meeting of the two leaders lowered the temperature". We suggest the following definition. The medium is an  information related to politics, its "internal energy" is its seriousness (in the sense of possible consequences), and the process of equilibration is the passage of the information by gossip.

\end{document}